%Paper: hep-th/9404077
%From: konstant@uful07.phys.ufl.edu (Konstantinos Anagnostopoulos)
%Date: Wed, 13 Apr 1994 17:57:50 -0400

\input phyzzx
\PHYSREV
\hoffset=0.3in
\voffset=-1pt
\baselineskip = 14pt \lineskiplimit = 1pt
\frontpagetrue
\rightline {Cincinnati preprint Dec.1993}
\medskip
\titlestyle{\seventeenrm Phase Transition in Rigid String Coupled to
Kalb-Ramond Fields}
\vskip.5in
\medskip
\centerline {\caps M. Awada and D. Zoller\footnote*{\rm E-Mail address:
moustafa@physunc.phy.uc.edu}}
\centerline {Physics Department}
\centerline {\it University of Cincinnati,Cincinnati, OH-45221}
\bigskip
\centerline {\bf Abstract}
\bigskip
Polyakov has argued that the QCD string should have long range order.
We show that a phase transition does exist in a generalization of
string theory characterized by the addition of the curvature of
the world sheet (rigidity) and the long range Kalb-Ramond interactions
to the Nambu-Goto action.  Although rigid strings coupled to long
range interactions exhibit the typical pathologies of higher derivative
theories at the classical level, we comment based on previous results
in rigid paths coupled to long range Coulomb interactions that both
phases of the quantum theory are free of ghosts and tachyons.

\eject

Polykov [1] has argued that the string theory appropriate to QCD should
be one with long range correlations of the unit normal. The Nambu-Goto
(NG) string theory is not the correct candidate for large N QCD as it
disagrees with it at short distances.  The NG string does not give
rise to the parton-like behaviour observed in deep inelastic scattering
at very high energies.  The observed scattering amplitudes have a power
fall-off behaviour contrary to the exponential fall-off behaviour of
the NG string scattering amplitudes at short distances.  The absence of
scale and the power law behaviour at short distances suggest that
the QCD string must have long range order at very high energies.
Pursuing this end, Polyakov considered modifying the Nambu action
by the renormalizable scale invariant curvature squared term
(rigid strings).  The theory closely resembles the two dimensional
sigma model where the unit normals correspond to the sigma fields.
In the large N approximation, there is no phase transition [2].
Polyakov suggested adding a topological term to produce a phase
transition to a region of long range order.  In this note,
we couple rigid strings instead to long range Kalb-Ramond fields.
Since spin systems in two dimensions may exhibit a phase
transition with the inclusion of long range interactions, it is
natural to conjecture likewise for rigid strings
with long range Kalb-Ramond fields.   Here we prove there is
indeed a phase transition to a region of long range order in
the large N approximation. Such a theory may therefore be
relevant to QCD.

Higher derivative theories are typically pathological and we
must address the consistency of the theory.  The rigid string
without long range interactions is consistent because the
coupling constant of the curvature is not constant: the
running coupling kills the curvature term in the
infared limit and the theory has linear Regge behaviour
with no ghosts.  Although we have not addressed  the issue
of ghosts in the rigid string with Kalb-Ramond fields,
we have extensively investigated the analog for
point particles [3]. This theory generalizes Feynman's first
quantized description of QED by adding the curvature of the
world line to the action.  This theory has two phases:
A disordered phase whose large distance scale is
essentially ordinary QED and the other is a new phase of QED
with long range correlations of the unit tangents to the
world lines.  We have explicitly shown in the large N limit
that both phases are free of ghosts
and tachyons by calculating the spacetime propagator.
We found that the ghosts are of order of the cut-off scale
and in the absence of fine tuning of the coupling constants
they decouple from the mass spectrum of the theory.
Thus, due to the phase transition, the quantum fluctuations
of the classical theory prevent one from taking the
classical limit and inheriting the
problems of the classical theory.  We believe that the same mechanism
would apply for our model of rigid strings coupled to the Kalb-Ramond
fields.  We will investigate this in a future article.

 The gauge fixed action of the rigid string [1] coupled to the rank
two antisymmetric Kalb-Ramond tensor field $\phi$ is [4]:
$$I_{gauge-fixed} = \mu_{0}\int d^2\xi\rho + {1\over 2t}\int d^2\xi
[\rho^{-1}(\partial^{2}x)^{2} + \lambda^{ab}(\partial_{a}x
\partial_{b}x - \rho\delta_{ab})] + I_{Kalb-Ramond}\eqno{(1 a)}$$
where
$$I_{K-R}= e_{0}\int d^2\xi \epsilon^{ab}\partial_{a}x^{\mu}
\partial_{b}x^{\nu}\phi_{\mu\nu} + {1\over 12}\int d^4x
F_{\mu\nu\rho}F^{\mu\nu\rho}\  .\eqno{(1 b)}$$
where $e_{0}$ is a coupling constant of dimension $length^{-1}$,
t is the curvature coupling constant which is dimensionless and F
is the abelian field strength of $\phi$.  The integration of the
$\phi$ field is Gaussian.  We obtain the following interacting
long range Coulomb-like term that modifies the rigid string:
$${1\over 2t}\int\int d^2\xi d^2\xi' \sigma^{\mu\nu}(\xi)
\sigma_{\mu\nu}(\xi')V(|x-x'|, a)\eqno{(1 c)}$$
where V is the  analog of the long range Coulomb potential:
$$ V(|x-x'|,a)= {2g\over \pi}{1\over |x(\xi)-x(\xi')|^{2}
+a^{2}\rho}\   .\eqno{(1 d)}$$
where $\sigma^{\mu\nu}(\xi)= \epsilon^{ab}\partial_{a}x^{\mu}
\partial_{b}x^{\nu}$.
We have introduced the cut-off "a" to avoid the singularity at
$\xi=\xi'$ and define $ g=t.\alpha_{Coulomb}=t.{e^{2}\over 4\pi}$
which has dimension of $length^{-2}$.  The partition function is
$$ Z =\int D\lambda D\rho Dx exp(-I_{eff})\  .\eqno{(2)}$$
where the effective action $I_{eff}$ is (1a) and (1c).

{\bf II-Large D analysis, absence of the Kalb-Ramond Coulomb
interactions}

The effective action is obtained by integrating over
$x^{\nu}, \nu=1,...D$
we have:
$$ {I_{0}}_{eff} = {1\over 2t}[\int d^2\xi (\lambda^{ab}
(-\rho\delta_{ab}) + 2t\mu_{0}\rho) + tDtrln A]\eqno{(3)}$$
where A is the operator
$$ A = \partial^{2}\rho^{-1}\partial^{2} -\partial_{a}
\lambda^{ab}\partial_{b}\  .\eqno{(4)}$$
In the large D limit the stationary point equations resulting
from varying $\lambda$ and $\rho$ respectively are:
$$\rho={tD\over 2}trG\eqno {(5 a)}$$
$$2t\mu_{0}-\lambda^{ab}\delta_{ab}=tDtr(\rho^{-2}
(-\partial^{2}G))\eqno {(5 b)}$$
where the world sheet Green's function is defined by:
$$ G(\xi,\xi') = <\xi|(-\partial^{2})A^{-1}|\xi'>\eqno{(6)}$$
The stationary points are:
$$ \rho(\xi) = \rho^{*},~~~~~~~~\lambda^{ab}=\lambda^*\delta^{ab}
\eqno{(7)}$$
where $\rho^{*}$ and $\lambda^*$ are constants.  Thus eq.(5a)
becomes the mass gap equation:
$$ 1={Dt\over 2}\int{d^2p\over (2\pi)^{2}}{1\over p^2+m^2}
\eqno{(8)}$$
where we define the mass
$$m^{2}=\rho^*\lambda^*\eqno{(9)}$$
this yields the mass gap equation,
$$ m = \Lambda e^{-{4\pi\over Dt}}\eqno{(10 a)}$$
where $\Lambda={1\over a}$ is an U.V. cut-off and m is now
the mass associated with the propagator:
$$ <\partial_{a} x^{\mu}(p)\partial_{a} x^{\nu}(-p)> =
{Dt\over 2}{\delta^{{\mu}{\nu}}\over p^2+m^2}
\  .\eqno{(10 b)}$$
On the other hand eq(5b) yields the string tension
renormalization condition:
$$\mu_{0}= {D\over 8\pi}{\Lambda^{2}\over \rho^*}
\eqno{(11)}$$
eq. (10a) agrees exactly with the one loop result.

{\bf III-Phase transition in the presence of Kalb-Ramond Coulomb
interactions}

The integration is no longer Gaussian.  Thus we consider
$$x^{\nu}(\xi) = x^{\nu}_{0}(\xi) + x^{\nu}_{1}(\xi)$$
and expand the Coulomb term (1c,d) to quadratic order in
$x^{\nu}_{1}(\xi)$ about the background straight line $x_{0}$.
The x-integration is now Gaussian and the
new effective action $S_{eff}$ is given by (3) and (4)
however with a new operator
A that includes the Coulomb potential contributions.
Using the stationary solutions (7) we obtain:
$$ trln A_{new} = \int{d^2p\over (2\pi)^2}ln[p^{4}
+ p^{2}m^{2} + p^{2}V_{0}(p) + V_{1}(p)]\eqno{(12)}$$
where
$$ V_{0}(p) = {4g\over \pi}\int d^2\xi
{e^{ip.\xi}\over \xi^{2}+a^{2}}= 8gK_{0}(a|p|)\eqno{(13 a)}$$
$$V_{1}(p)={8g\over \pi}\int d^2\xi
{[e^{ip.\xi}-1]\over (\xi^{2}+a^{2})^{2}}= {8g\over a^2}
(aK_{1}(a|p|)-1)\eqno{(13 b)}$$
where $K_{n}(z), n=0,1,..$ is the Bessel function of the third
kind and
$$K_{1}(z) := -{d\over dz}K_{0}(z)\  .\eqno{(14)}$$
The new mass gap equation given by (5a),(6) and (12) that
generalizes (8) is:
$$ 1 ={Dt\over 2}\int{d^2p\over (2\pi)^2}
{p^{2}\over p^{2}(p^{2}+m^{2}) + p^{2}V_{0}(p)
+ V_{1}(p)}\eqno{(15)}$$
The critical line is defined by eq.(15) at $m=0$:
$$ 1 ={Dt\over 4\pi}\int_{0}^{1} dy{y^{3}\over y^{4}+
\eta(y^{2}K_{0}(y)+yK_{1}(y)-1)}\eqno{(16)}$$
where $\eta=8ga^{2}$ is a dimensionless coupling
constant. We have made the change of
variable y=ap and introduced the U.V cut-off
$\Lambda ={1\over a}:=\Lambda_{0}$ \footnote*{ In fact there
exist an $\eta^{*}$ for which any choice of $\Lambda$a=c leads
to phase transition as long as $\eta<\eta^{*}$.  We choose
$\Lambda=\Lambda_{Planck}$, therefore $c\geq 1$.} .
It is remarkable that eq.(16) is finite except at g=0
(absence of Coulomb interactions). In fact after tedious
calculations one can prove that
$$ lim_{y\rightarrow 0}
[{y^{3}\over y^{4}+\eta(y^{2}K_{0}(y)+yK_{1}(y)-1)}]=0$$
thus having infra-red finiteness.

The critical curve distinguishing the two phases in
the $(t,\eta)$ plane is shown in Fig.1.  The order
parameter of the theory is the mass gap equation
(15) where m is the parameter that distinguishes
that two phases.  In the disordered phase $ m >0$,
while in the ordered phase it is straightforward to
show that m =0.   In the disordered phase the
coupling constants t and g are completely fixed by
dimensional transmutation in terms of the cut-off
$\Lambda$ and m.

In conclusion, we have shown that the rigid string
coupled to long range Kalb-Ramond fields has a phase
transition in the ${1\over N}$ approximation.
The new phase is characterized by long range order
 with absence of scale and may describe QCD strings.
Further questions for future investigations include
the issue of mass spectrum of the theory
and the QCD loop equation.

{\bf Acknowledgment}

 We are very grateful to prof. A. Polyakov for his
constant encourgement and extensive support and
especially for suggesting his conjecture that a
higher derivative regulated quantum theory does
not necessarily have the pathologies of the classical
theory. One of us M.A is thankful to him for long
and patient conversations explaining the decoupling
mechanism of ghosts and its connection with critical
phenomena.  We are also grateful to Prof. Y. Nambu
for his extensive support, encourgement
and long discussions over the last two years with out
which we could not have gone far in our investigations.
We also Thank F. Mansouri for constructive discussions
and suggestions, J. Clark and D.Seifu for the graphical
representation of the critical line and using numerical
methods for locating the poles of the Green's function.
We finally thank P.Ramond for encourgement and support
and for suggesting the study of the Kalb-Ramond theory
in connection with the rigidity of the string.  This
generalizes a similar result we found in our model of
rigid QED [3].

{\bf References}

\item {[1]} A. Polyakov, Nucl. Phys. B268 (1986) 406
; A. Polyakov,  Gauge fields, and Strings,
Vol.3, harwood academic publishers
\item {[2]} Alfonso and M. Espero,
Nucl. Phys. B278 (1987) 325
\item {[3]} M. Awada and D. Zoller,
Cincinnati preprint Dec.1993-1, to appear in Phys. Lett.B,
Phys.Lett B299 (1993) 151, also see the
detailed version: Cincinnati preprint Jan. 1993-105.
To appear in Int. Journal. Of Physics A. M.Awada, M.Ma,
and D.Zoller Mod.phys. Lett A8,(1993), 2585
\item {[4]} M. Kalb, and P. Ramond
Phys. Rev. D Vol.9 (1974) 2237
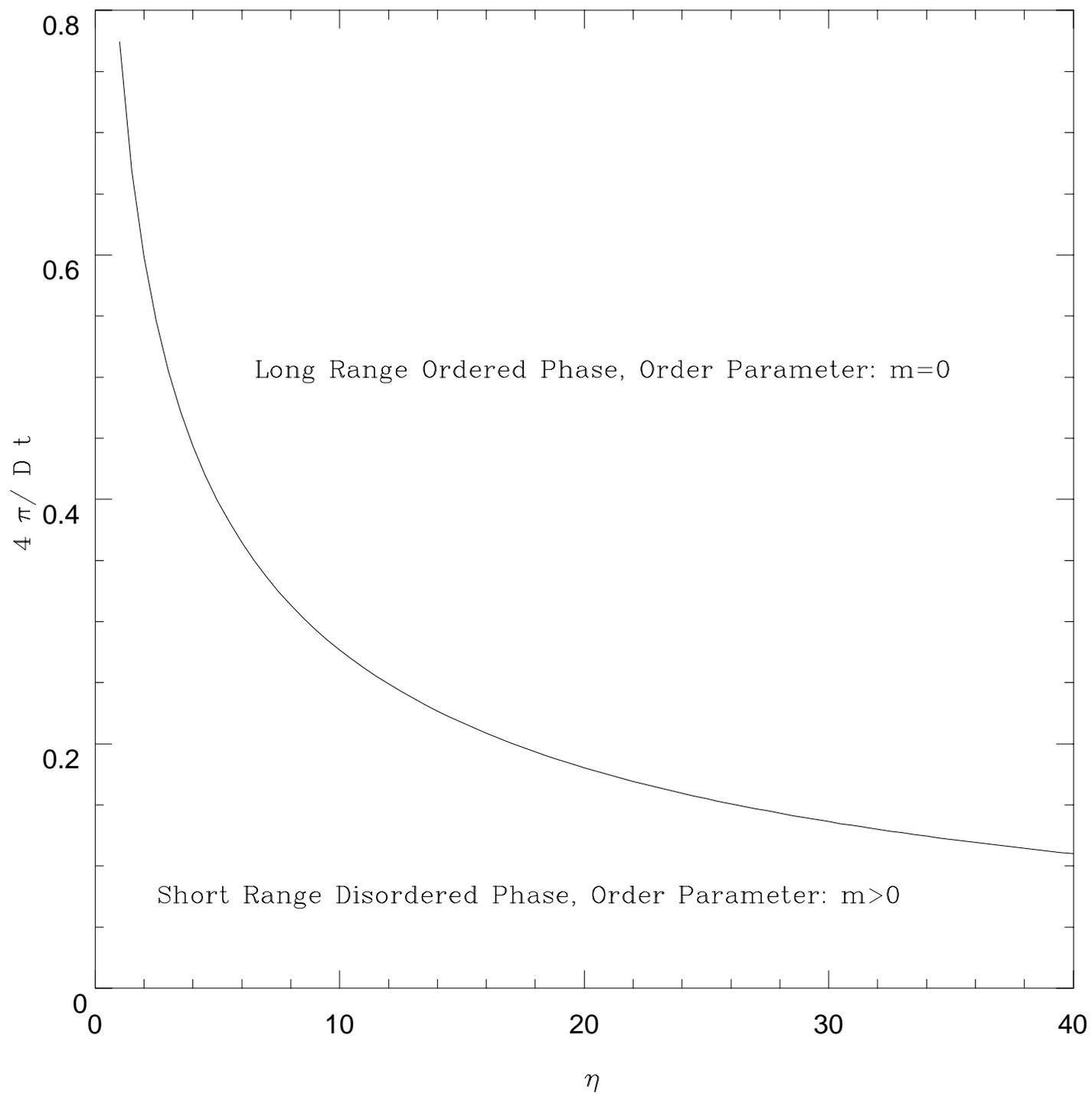
\end